\documentclass[letter,bibyear]{aa}
\usepackage{graphicx}
\usepackage[]{natbib}
\usepackage{txfonts}
\usepackage{graphicx}
\usepackage{amsmath}

\usepackage{amssymb}
\usepackage{orcidlink}
\usepackage{upgreek}
\usepackage{ulem}
\usepackage{xcolor}

\newcommand{\St}{\mathrm{St}}
\defcitealias{Michoulier2024_PorosityEvolution}{Paper~I}

\begin{document} 

   \title{The CO snow line favours strong clumping by the streaming instability in protoplanetary discs with porous grains}
   \titlerunning{The CO snow line with porous grains favours strong clumping by the SI}

   \author{Jean-François Gonzalez\inst{1}
          \and
          Stéphane Michoulier\inst{1}
         }

   \institute{Universite Claude Bernard Lyon 1, CRAL UMR5574, ENS de Lyon, CNRS, Villeurbanne 69622, France\\
              \email{jean-francois.gonzalez@ens-lyon.fr}
            }

   \date{Received 30 October 2025 / Accepted 24 January 2026}

  \abstract{The radial drift and fragmentation of small dust grains in protoplanetary discs impedes their growth past centimetre sizes. Several mechanisms have been proposed to overcome these planet formation barriers, such as dust porosity or the streaming instability (SI), which is today regarded as the most promising mechanism to form planetesimals.}
  {Here, we examine whether the conditions for the SI to lead to strong clumping, the first step in planetesimal formation, are realised in protoplanetary discs containing porous grains.}
  {We used results from previous simulations of the evolution of porous grains subjected to growth, fragmentation, compaction and bouncing in protoplanetary discs. In the ensuing disc structures, we determined the regions where the dust-to-gas ratio exceeds the critical value for strong clumping found in simulations of the SI including external turbulence.}
  {We find that the conditions for strong clumping are met within the first hundred thousand years in large regions of protoplanetary discs containing porous grains, provided that the CO snow line is taken into account. If the CO snow line is neglected, the conditions are met only very close to the inner disc edge early on, or over large areas well after 200,000~yr.}
  {}
 

   \maketitle
   \nolinenumbers

\section{Introduction}
\label{Sec:Intro}

In the core accretion paradigm \citep{Pollack1996,Alibert2005}, planet formation is thought to proceed from the growth of sub-micrometre-sized dust particles into larger and larger solids. Small grains easily stick upon collision up to cm-sized pebbles thanks to van der Waals forces and km-sized planetesimals can rapidly coalesce into planetary cores, and ultimately planets, due to gravity. However, the stage from pebbles to planetesimals is hindered by several processes, known as barriers to planet formation.
As dust grains grow, their relative velocities increase \citep{Weidenschilling1993}. Above a threshold velocity $v_\mathrm{frag}$, grains shatter because of excess kinetic energy, this is the fragmentation barrier \citep{Blum2008}. Near this limit, grains only bounce without growing, this is the bouncing barrier \citep{Zsom2010}. In both cases, occurring near pebble sizes, grains cannot contribute to the formation of larger bodies.
Furthermore, solid bodies immersed in a gas disc are subject to aerodynamic drag, which causes them to drift inwards \citep{Whipple1972}. Its importance is quantified by the Stokes number St, defined as the ratio of the drag stopping time to the particle's orbital time. Small grains, with $\St\ll1$, are strongly coupled to the gas and mostly follow its motion. Large solids, with $\St\gg1$, are largely decoupled and orbit the star independently of the gas. Particles in the intermediate regime ($\St\sim1$) are also mostly of pebble sizes and undergo the most efficient drift, often reaching the star before growing further. This is the radial-drift barrier \citep{Adachi1976,Weidenschilling1977}.

Solving these problems requires mechanisms allowing particles to reach large St values and decouple from the gas before experiencing significant drift. This can be achieved by either accelerating dust growth or slowing down radial drift. Taking into account grain porosity explores the former. Porous grains are observed in Solar System asteroids \citep{Consolmagno2008} or comets \citep{Guttler2019}, as well as in protoplanetary discs \citep[e.g.][]{Kataoka2016,Pinte2019,Tazaki2023}, where \citet{Liu2024} additionally showed that taking into account porous dust opacities may alleviate the dust mass budget problem. Laboratory and numerical experiments have also shown that collisions of small grains regularly form porous aggregates \citep{Blum2000}. For a given mass, porous grains have a larger collisional cross-section and thus grow faster than their compact counterparts. While the latter mostly stay in the Epstein drag regime, in which $\St\propto s$, with $s$ the grain size, the former can reach the Stokes regime, where $\St\propto s^2$ and thus attain large St values and decouple sooner \citep{Ormel2007,Okuzumi2012,Garcia2020}. Recently, we have shown in numerical simulations of grain evolution that taking into account porosity and compaction during fragmentation is necessary to reproduce observed grain properties in discs \citep[][hereafter Paper~I]{Michoulier2024_PorosityEvolution}.

A second class of solutions involves slowing down or stopping radial drift. Dust particles drift towards pressure maxima in the disc \citep{Whipple1972}, accumulating there in so-called dust traps. Their confinement keeps their relative velocities lower than the bouncing and fragmentation thresholds, allowing them to grow unhindered. Various phenomena can cause such traps and have been extensively studied, e.g.\ vortices \citep{Meheut2012}, planet gap edges \citep{Paardekooper2004,Fouchet2007,Fouchet2010}, snow lines \citep{Kretke2007,Vericel2020}, or self-induced dust traps \citep{Gonzalez2015,Gonzalez2017,Vericel2021}.

Finally, instabilities can lead to the direct formation of planetesimals. The most famous one is the streaming instability (SI), discovered by \citet{Youdin2005}. The SI is a hydrodynamic instability caused by the relative drift of gas and dust, which grows faster when the local dust-to-gas ratio $\varepsilon=\rho_\mathrm{d}/\rho_\mathrm{g}$, where $\rho_\mathrm{d}$ and $\rho_\mathrm{g}$ are the volume densities of dust and gas, respectively, is high (typically 0.1--1 or more) and the dust coupling to the gas is marginal ($\St\lesssim1$). Physical interpretations for the SI have been detailed by \citet{Squire2020,Magnan2024}. The linear growth of the instability has been studied analytically \citep{Youdin2005,Youdin2007,Jacquet2011,Auffinger2018,Jaupart2020} and leads to enhancements in the particle density. Its non-linear growth has been investigated via shearing-box simulations \citep{Johansen2007_Nonlinear_SI,Schafer2017,Yang2017} and results in the spontaneous concentration of dust into clumps. If their self-gravity is large enough, these clumps can gravitationally collapse and directly form planetesimals \citep{Johansen2007_Planetesimal_formation}. The SI additionally allows to explain the formation and demographics of rocky bodies in the Solar System \citep{Li2019}. It is currently considered the leading mechanism to form planetesimals in protoplanetary discs \citep{Lesur2023}.

As a result, the conditions leading to strong particle clumping have been actively investigated \citep{Johansen2009,Bai2010,Carrera2015,Yang2017}: the metallicity $Z=\Sigma_\mathrm{d}/\Sigma_\mathrm{g}$, defined as the ratio of the dust-to-gas column densities\footnote{Some authors use $Z=\Sigma_\mathrm{d}/(\Sigma_\mathrm{g}+\Sigma_\mathrm{d})$, which is numerically very close.}, needs to exceed a critical value, found to be $Z_\mathrm{crit}\sim0.02$ for $\St\sim0.1$, and higher for smaller or larger $\St$. With higher-resolution simulations, \citet{Li2021,Lim2025} found reduced values of $Z_\mathrm{crit}$, down to 0.004 for the optimal $\St$. However, including particle self-gravity and external turbulence, \citep{Lim2024} found that $Z_\mathrm{crit}$ is increased by up to one order of magnitude. In any case, whether these criteria are met in real discs has not yet been verified.

In this Letter, we assessed the viability of the SI in protoplanetary discs with porous dust grains using the simulations of porous dust evolution presented in \citetalias{Michoulier2024_PorosityEvolution}. The strong clumping criteria are expressed in terms of $\St$, which depends both on the grains size $s$ and their filling factor $\phi$ \citep[see][]{Garcia2020}. Our simulations, providing both $s$ and $\phi$, allow us to directly compute $\St$ anywhere in the disc, thus lifting the degeneracy between $s$ and $\phi$. We present our methods in Sect.~\ref{Sec:Methods}, our results in Sect.~\ref{Sec:Results}, and discuss them in Sect.~\ref{Sec:Discussion}. We conclude in Sect.~\ref{Sec:Conclusion}.

\section{Methods}
\label{Sec:Methods}

In \citetalias{Michoulier2024_PorosityEvolution}, we presented our grain porosity evolution model, applied it to a suite of simulations of a gas and dust protoplanetary disc with the 3D Smoothed Particle Hydrodynamics (SPH) code \textsc{Phantom} \citep{Price2018}, and extensively discussed their results. Here, we exploit a subset of those simulations that best reproduce observations: they include compaction during fragmentation (see \citetalias{Michoulier2024_PorosityEvolution} for details). Focussing on porous silicate grains, we selected two cases: without (simulation GBFc-Si-a02-Vf20) and with (GBFcS-Si-a02-Vf205) the CO snow line.

We recall the simulation setup: the $0.01~$M$_\odot$ disc, with a radial extent from 10 to 400~au, a surface density profile \mbox{$\Sigma\propto (r/r_0)^{-0.75}$} and a temperature profile $T\propto (r/r_0)^{-0.5}$, orbits a 1~M$_\odot$ star. Its aspect ratio is $(H/R)_0=0.0895$ at $r_0=100$~au, and we used a turbulent viscosity parameter \citep{SS73} $\alpha=5\times10^{-3}$. Grains of intrinsic density $\rho_\mathrm{s}=2\,700$~kg\,m$^{-3}$ and fragmentation threshold $v_\mathrm{frag}=20$~m\,s$^{-1}$ are initially $\mu$m-sized with uniform dust-to-gas ratio $\varepsilon_0=0.01$. Beyond the CO snow line at $T=20$~K ($r\gtrsim100$~au in our disc), the CO ice coating is assumed to lower $v_\mathrm{frag}$ to 5~m\,s$^{-1}$. The simulations comprised 1.2 million SPH particles in the `dust-as-mixture' formalism and were evolved for 300\,000~yr. The reader is referred to \citetalias{Michoulier2024_PorosityEvolution} for more information on the simulation suite and setup.

The results of our global 3D simulations are recalled in Appendix~\ref{App:ResultsSummary}. As is the case with all SPH simulations to date \citep{David-Cleris2024}, they lacked the resolution needed to capture the SI. Instead, in this Letter, we investigated whether strong clumping by the SI can occur in our simulated disc by examining if the required conditions were met in our two cases. Unlike most authors, we did not use the critical metallicity $Z_\mathrm{crit}$, which involves vertically integrated surface densities. Instead, we preferred to rely on the local dust-to-gas ratio $\varepsilon$ computed from volume densities. Indeed, we thought it to be more relevant to study conditions at different altitudes above the midplane in the presence of dust settling. We adopted the critical value found by \citet{Lim2024} and given by
\begin{equation}
  \log\varepsilon_\mathrm{crit} = 0.42\,(\log\St)^2 + 0.72 \log\St + 0.37,
  \label{Eq:eps_crit}
\end{equation}
which they found to be mostly insensitive to the levels of turbulence they studied, contrary to $Z_\mathrm{crit}$. We considered that strong clumping occurred in the disc regions where $\varepsilon>\varepsilon_\mathrm{crit}$.

\section{Results}
\label{Sec:Results}

\begin{figure*}
  \centering
  \includegraphics[width=\textwidth,clip]{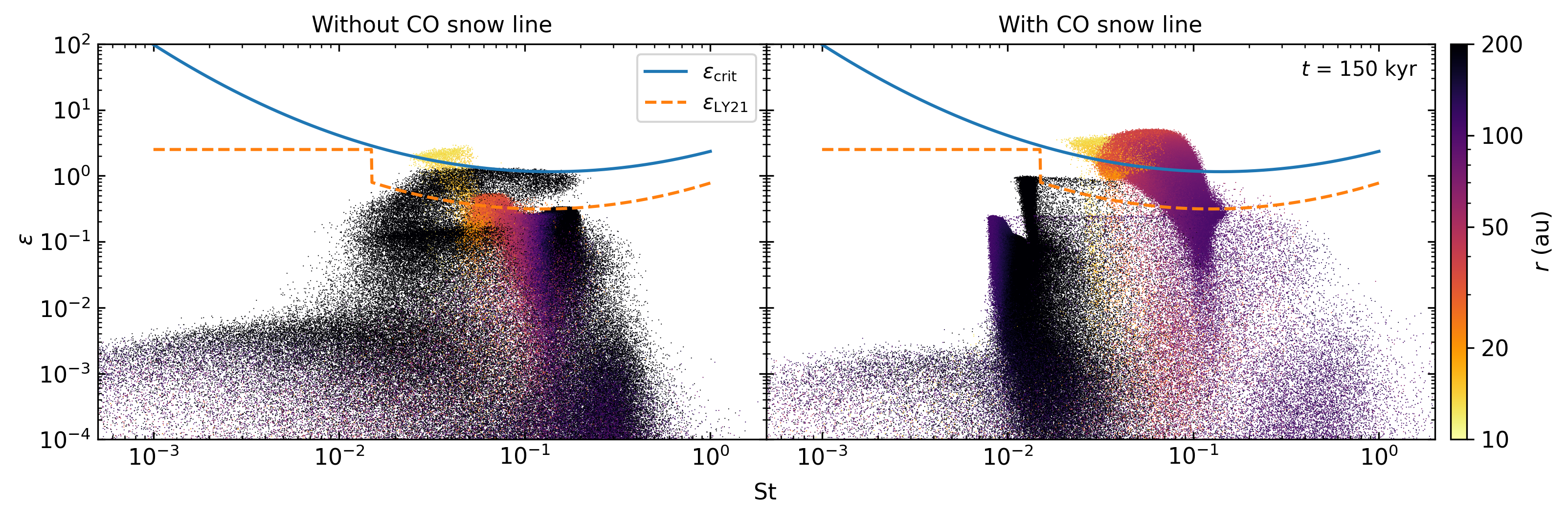}
  \caption{Positions of the SPH particles in the St-$\varepsilon$ plane, coloured by their radial distance to the star $r$, at $t=150,000$~yr in simulations without (\textit{left}) and with (\textit{right}) the CO snow line. The solid and dashed lines show the critical dust-to-gas ratios $\varepsilon_\mathrm{crit}$ from \citet{Lim2024} and $\varepsilon_\mathrm{LY21}$ from \citet{Li2021}, respectively.}
  \label{Fig:SI_trig_cond}
\end{figure*}

We examined the conditions for strong clumping by the SI in Fig.~\ref{Fig:SI_trig_cond}, which displays the locations of the SPH particles in the St-$\varepsilon$ plane at $t=150$~kyr. Their colour represents their radial distance to the star. The solid line plots $\varepsilon_\mathrm{crit}$ from Eq.~\ref{Eq:eps_crit}. We also show for reference (dashed line) the critical curve $\varepsilon_\mathrm{LY21}$ from \citet{Li2021}, which is widely used but neglects external turbulence. In the disc without CO snow line (left), the conditions for strong clumping are met only in a small fraction of the disc, at its inner edge (yellow points). With the CO snow line (right), however, the disc areas with strong clumping extend from the disc inner edge out to its intermediate regions (purple points).

\begin{figure*}
  \centering
  \includegraphics[width=\textwidth,clip]{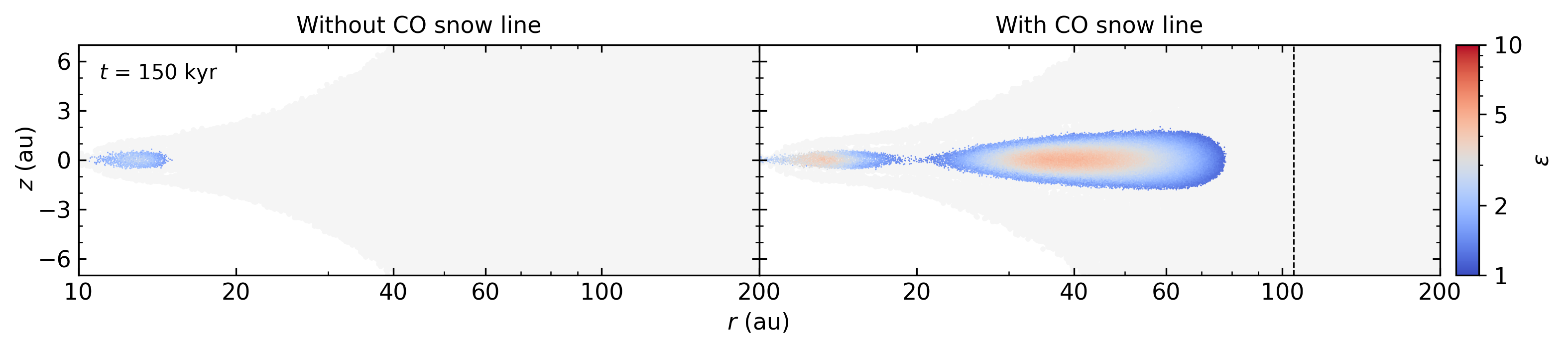}
  \caption{Map of the dust-to-gas ratio $\varepsilon$ where it exceeds $\varepsilon_\mathrm{crit}$ in the meridian plane $(r,z)$ at $t=150,000$~yr in simulations without (\textit{left}) and with the CO snow line, marked by the vertical dashed line (\textit{right}). The extent of the gas disc is shown in light grey.}
  \label{Fig:SI_location}
\end{figure*}

The precise locations of the disc regions where strong clumping can occur are better viewed in meridian plane maps of the areas where $\varepsilon>\varepsilon_\mathrm{crit}$, shown in Fig.~\ref{Fig:SI_location}, again at $t=150$~kyr. Without CO snow line (left), the strong-clumping region is limited to the inner disc interior to $\sim$15~au. In the disc with the CO snow line (right), it spreads from the inner edge out to $\sim$80~au. It includes two zones of larger dust-to-gas ratio with peak values as high as $\sim$5, centred around $\sim$15 and $\sim$40~au. The outer one reaches a couple of au above the midplane.

The time evolution of the strong-clumping regions is shown in Appendix~\ref{App:TimeEvol}. The inner one forms early on, at 20~kyr, interior to 15~au in both cases. With the CO snow line, an outer strong-clumping region forms near 40~au at 70~kyr and both regions expand and merge at 150~kyr.
At later times, without CO snow line the strong-clumping region at the disc inner edge shrinks and disappears while a second region starts to grow around 35~au. By 300~kyr, the latter extends from $\sim$25 to $\sim$90~au with peak values of $\varepsilon\gtrsim5$. In contrast, with the CO snow line the strong-clumping region evolves only slightly in the radial direction but thickens, with a dust-to-gas ratio that continues to increase.

\section{Discussion}
\label{Sec:Discussion}

Strong clumping by the SI requires $\varepsilon$ to become larger than $\varepsilon_\mathrm{crit}$. Dust trapping is an efficient mechanism to increase $\varepsilon$ (see Sect.~\ref{Sec:Intro}). Dust from infalling envelopes can also lead to dust-enriched regions in the embedded disc phase \citep{Cridland2022}.
In our discs, the increase in $\varepsilon$ is due to a combination of the vertical settling and radial drift of growing dust. When the CO snow line is not included, grains drift fast (because of their large St, see Appendix~\ref{App:ResultsSummary}) from the outer disc until St drops to a few percent, slow down and pile up interior to $\sim$15~au, where $\varepsilon$ exceeds unity.
Later on, the ongoing grain compaction lowers their St at progressively larger distances from the star, resulting in the appearance of a second strong-clumping region at intermediate distances.
When it is present, the CO snow line acts as a barrier that slowly feeds the disc interior to it, preventing grains to grow too fast and limiting their St to at most 0.1. They drift more slowly and accumulate where St decreases, farther out than in the previous case, resulting early on in a region of enhanced $\varepsilon$ tens of au from the star. Over time, this region grows slowly, merges with the inner one, and becomes more dust-enriched.

Planetesimal formation additionally needs large areas where strong clumping can occur. Indeed, if these areas are too narrow, drifting grains can cross them faster than the SI can grow \citep{Carrera2022}. This can be an issue at early stages without CO snow line. However, the more realistic case, including all relevant processes (grain porosity, growth, fragmentation with compaction, bouncing, and the CO snow line), is also the most favourable, with an early development of extended areas conducive to planetesimal formation. This is supported by a crude estimate of their available mass reservoir (see Appendix~\ref{App:MassReservoir}).

Besides those inherent to our dust evolution model and extensively discussed in \citetalias{Michoulier2024_PorosityEvolution}, the main caveat in this work is that the condition for strong clumping that we used, $\varepsilon>\varepsilon_\mathrm{crit}$, was derived for the monodisperse SI (mSI), i.e.\ considering only one St value at a time. Yet, our simulations of dust evolution include a wide range of grains sizes, and therefore of St numbers. Studies of the SI taking into account a particle size distribution (polydisperse SI, PSI) find that growth rates are much lower than for the mSI \citep{Krapp2019,Paardekooper2020,Matthijsse2025}. However, both the linear growth for size distributions peaking at the largest grains if $\varepsilon\gtrsim1$ \citep{McNally2021} and non-linear saturation states in the fast-growth regime \citep{Yang2021} are similar to those of the mSI.
When coagulation is taken into account, dust growth and the SI can assist each other in a positive feedback loop \citep{Tominaga2023,Carrera2025}, which can help the dust-to-gas ratio to rise above critical values.
The situation becomes more complex in the presence of magnetic fields: \citet{Lin2022} found that the mSI can be stabilised by magnetic perturbations, in particular when $\varepsilon$ is low. However, exploratory work of the magnetised PSI \citep{David-Cleris2024} has suggested that the presence of a magnetic field makes the SI more robust to polydisperse dampening.
In any case, simulations aiming to determine the criteria for strong clumping by the PSI would be very valuable to assess more realistically where planetesimals can form.

\section{Conclusion}
\label{Sec:Conclusion}

We used simulations of the dynamics of porous dust experiencing growth, fragmentation, bouncing and compaction in protoplanetary discs that we ran in \citetalias{Michoulier2024_PorosityEvolution} to explore if and where the conditions for strong clumping by the SI are met. 
We found that the evolution of porous grains results in a strong-clumping region limited to the disc interior to 15~au for $\sim$200~kyr. It is only when the CO snow line, which helps to retain more large grains in the disc, is taken into account that the situation brightens, with the early appearance of strong clumping extending over a large area out to a few tens of au. The presence of the CO snow line thus appears to be key to promote planetesimal formation in discs with porous grains. While the physical mechanism is different, its importance can be related to that of the water snow line much closer in \citep{Drazkowska2017}.

In future work, we envision performing simulations of the SI in dusty discs together with porous grain evolution using the newly-released, GPU-accelerated \textsc{Shamrock} code \citep{David-Cleris2025}.

\begin{acknowledgements}
We thank Antoine Alaguero, Daniel Carrera, Timothée David-Cléris and Guillaume Laibe
for fruitful discussions, as well as the anonymous referee, whose report helped improve this Letter.
The authors acknowledge funding from ANR (Agence Nationale de la Recherche) of France under contract number ANR-16-CE31-0013 (Planet-Forming-Disks) and thank the LABEX Lyon Institute of Origins (ANR-10-LABX-0066) for its financial support within the Plan France 2030 of the French government operated by the ANR.
This research was partially supported by the Programme National de Physique Stellaire and the Programme National de Planétologie of CNRS (Centre National de la Recherche Scientifique)/INSU (Institut National des Sciences de l’Univers), France.
This project has received funding from the European Union's Horizon 2020 research and innovation programme under the Marie Sk\l{}odowska-Curie grant agreements No 210021 and No 823823 (DUSTBUSTERS).
We gratefully acknowledge support from the CBPsmn (PSMN, Pôle Scientifique de Modélisation Numérique) of the ENS de Lyon for the computing resources. The platform operates the SIDUS solution \citep{Quemener2013} developed by Emmanuel Quemener.
Figures were made with the Python library \texttt{matplotlib} \citep{Hunter2007}.
The data necessary to recreate the \textsc{Phantom} simulations and to reproduce the figures are available in \citet{Michoulier2024_PorosityEvolution_dataset}.
\end{acknowledgements}

\bibliographystyle{aa}
\bibliography{SI_triggering}

\begin{thebibliography}{65}
\expandafter\ifx\csname natexlab\endcsname\relax\def\natexlab#1{#1}\fi

\bibitem[{{Adachi} {et~al.}(1976){Adachi}, {Hayashi}, \&
  {Nakazawa}}]{Adachi1976}
{Adachi}, I., {Hayashi}, C., \& {Nakazawa}, K. 1976, Progress of Theoretical
  Physics, 56, 1756

\bibitem[{{Alibert} {et~al.}(2005){Alibert}, {Mordasini}, {Benz}, \&
  {Winisdoerffer}}]{Alibert2005}
{Alibert}, Y., {Mordasini}, C., {Benz}, W., \& {Winisdoerffer}, C. 2005, \aap,
  434, 343

\bibitem[{{Auffinger} \& {Laibe}(2018)}]{Auffinger2018}
{Auffinger}, J. \& {Laibe}, G. 2018, \mnras, 473, 796

\bibitem[{{Bai} \& {Stone}(2010)}]{Bai2010}
{Bai}, X.-N. \& {Stone}, J.~M. 2010, \apj, 722, 1437

\bibitem[{{Blum} \& {Wurm}(2000)}]{Blum2000}
{Blum}, J. \& {Wurm}, G. 2000, \icarus, 143, 138

\bibitem[{{Blum} \& {Wurm}(2008)}]{Blum2008}
{Blum}, J. \& {Wurm}, G. 2008, \araa, 46, 21

\bibitem[{{Carrera} {et~al.}(2015){Carrera}, {Johansen}, \&
  {Davies}}]{Carrera2015}
{Carrera}, D., {Johansen}, A., \& {Davies}, M.~B. 2015, \aap, 579, A43

\bibitem[{{Carrera} {et~al.}(2025){Carrera}, {Lim}, {Eriksson}, {Lyra}, \&
  {Simon}}]{Carrera2025}
{Carrera}, D., {Lim}, J., {Eriksson}, L. E.~J., {Lyra}, W., \& {Simon}, J.~B.
  2025, \aap, 696, L23

\bibitem[{{Carrera} \& {Simon}(2022)}]{Carrera2022}
{Carrera}, D. \& {Simon}, J.~B. 2022, \apjl, 933, L10

\bibitem[{{Consolmagno} {et~al.}(2008){Consolmagno}, {Britt}, \&
  {Macke}}]{Consolmagno2008}
{Consolmagno}, G., {Britt}, D., \& {Macke}, R. 2008, Chemie der Erde /
  Geochemistry, 68, 1

\bibitem[{{Cridland} {et~al.}(2022){Cridland}, {Rosotti}, {Tabone},
  {Tychoniec}, {McClure}, {Nazari}, \& {van Dishoeck}}]{Cridland2022}
{Cridland}, A.~J., {Rosotti}, G.~P., {Tabone}, B., {et~al.} 2022, \aap, 662,
  A90

\bibitem[{{David-Cl{\'e}ris}(2024)}]{David-Cleris2024}
{David-Cl{\'e}ris}, T. 2024, Phd thesis, {Universit{\'e} de Lyon}, available at
  https://theses.hal.science/tel-04816769

\bibitem[{{David-Cl{\'e}ris} {et~al.}(2025){David-Cl{\'e}ris}, {Laibe}, \&
  {Lapeyre}}]{David-Cleris2025}
{David-Cl{\'e}ris}, T., {Laibe}, G., \& {Lapeyre}, Y. 2025, \mnras, 539, 1

\bibitem[{{Dr{\k{a}}{\.z}kowska} \& {Alibert}(2017)}]{Drazkowska2017}
{Dr{\k{a}}{\.z}kowska}, J. \& {Alibert}, Y. 2017, \aap, 608, A92

\bibitem[{{Fouchet} {et~al.}(2010){Fouchet}, {Gonzalez}, \&
  {Maddison}}]{Fouchet2010}
{Fouchet}, L., {Gonzalez}, J.~F., \& {Maddison}, S.~T. 2010, \aap, 518, A16

\bibitem[{{Fouchet} {et~al.}(2007){Fouchet}, {Maddison}, {Gonzalez}, \&
  {Murray}}]{Fouchet2007}
{Fouchet}, L., {Maddison}, S.~T., {Gonzalez}, J.~F., \& {Murray}, J.~R. 2007,
  \aap, 474, 1037

\bibitem[{{Garcia} \& {Gonzalez}(2020)}]{Garcia2020}
{Garcia}, A. J.~L. \& {Gonzalez}, J.-F. 2020, \mnras, 493, 1788

\bibitem[{{Gonzalez} {et~al.}(2017){Gonzalez}, {Laibe}, \&
  {Maddison}}]{Gonzalez2017}
{Gonzalez}, J.~F., {Laibe}, G., \& {Maddison}, S.~T. 2017, \mnras, 467, 1984

\bibitem[{{Gonzalez} {et~al.}(2015){Gonzalez}, {Laibe}, {Maddison}, {Pinte}, \&
  {M{\'e}nard}}]{Gonzalez2015}
{Gonzalez}, J.~F., {Laibe}, G., {Maddison}, S.~T., {Pinte}, C., \&
  {M{\'e}nard}, F. 2015, \mnras, 454, L36

\bibitem[{{G{\"u}ttler} {et~al.}(2019){G{\"u}ttler}, {Mannel}, {Rotundi},
  {Merouane}, {Fulle}, {Bockel{\'e}e-Morvan}, {Lasue}, {Levasseur-Regourd},
  {Blum}, {Naletto}, {Sierks}, {Hilchenbach}, {Tubiana}, {Capaccioni},
  {Paquette}, {Flandes}, {Moreno}, {Agarwal}, {Bodewits}, {Bertini}, {Tozzi},
  {Hornung}, {Langevin}, {Kr{\"u}ger}, {Longobardo}, {Della Corte}, {T{\'o}th},
  {Filacchione}, {Ivanovski}, {Mottola}, \& {Rinaldi}}]{Guttler2019}
{G{\"u}ttler}, C., {Mannel}, T., {Rotundi}, A., {et~al.} 2019, \aap, 630, A24

\bibitem[{{Hunter,}(2007)}]{Hunter2007}
{Hunter,}, J.~D. 2007, Computing in Science \& Engineering, 9, 90

\bibitem[{{Jacquet} {et~al.}(2011){Jacquet}, {Balbus}, \&
  {Latter}}]{Jacquet2011}
{Jacquet}, E., {Balbus}, S., \& {Latter}, H. 2011, \mnras, 415, 3591

\bibitem[{{Jaupart} \& {Laibe}(2020)}]{Jaupart2020}
{Jaupart}, E. \& {Laibe}, G. 2020, \mnras, 492, 4591

\bibitem[{{Johansen} {et~al.}(2007){Johansen}, {Oishi}, {Mac Low}, {Klahr},
  {Henning}, \& {Youdin}}]{Johansen2007_Planetesimal_formation}
{Johansen}, A., {Oishi}, J.~S., {Mac Low}, M.-M., {et~al.} 2007, \nat, 448,
  1022

\bibitem[{{Johansen} \& {Youdin}(2007)}]{Johansen2007_Nonlinear_SI}
{Johansen}, A. \& {Youdin}, A. 2007, \apj, 662, 627

\bibitem[{{Johansen} {et~al.}(2009){Johansen}, {Youdin}, \& {Mac
  Low}}]{Johansen2009}
{Johansen}, A., {Youdin}, A., \& {Mac Low}, M.-M. 2009, \apjl, 704, L75

\bibitem[{{Kataoka} {et~al.}(2016){Kataoka}, {Muto}, {Momose}, {Tsukagoshi}, \&
  {Dullemond}}]{Kataoka2016}
{Kataoka}, A., {Muto}, T., {Momose}, M., {Tsukagoshi}, T., \& {Dullemond},
  C.~P. 2016, \apj, 820, 54

\bibitem[{{Krapp} {et~al.}(2019){Krapp}, {Ben{\'\i}tez-Llambay}, {Gressel}, \&
  {Pessah}}]{Krapp2019}
{Krapp}, L., {Ben{\'\i}tez-Llambay}, P., {Gressel}, O., \& {Pessah}, M.~E.
  2019, \apjl, 878, L30

\bibitem[{{Kretke} \& {Lin}(2007)}]{Kretke2007}
{Kretke}, K.~A. \& {Lin}, D.~N.~C. 2007, \apjl, 664, L55

\bibitem[{{Lesur} {et~al.}(2023){Lesur}, {Flock}, {Ercolano}, {Lin}, {Yang},
  {Barranco}, {Benitez-Llambay}, {Goodman}, {Johansen}, {Klahr}, {Laibe},
  {Lyra}, {Marcus}, {Nelson}, {Squire}, {Simon}, {Turner}, {Umurhan}, \&
  {Youdin}}]{Lesur2023}
{Lesur}, G., {Flock}, M., {Ercolano}, B., {et~al.} 2023, in Astronomical
  Society of the Pacific Conference Series, Vol. 534, Protostars and Planets
  VII, ed. S.~{Inutsuka}, Y.~{Aikawa}, T.~{Muto}, K.~{Tomida}, \& M.~{Tamura},
  465

\bibitem[{{Li} \& {Youdin}(2021)}]{Li2021}
{Li}, R. \& {Youdin}, A.~N. 2021, \apj, 919, 107

\bibitem[{{Li} {et~al.}(2019){Li}, {Youdin}, \& {Simon}}]{Li2019}
{Li}, R., {Youdin}, A.~N., \& {Simon}, J.~B. 2019, \apj, 885, 69

\bibitem[{{Lim} {et~al.}(2024){Lim}, {Simon}, {Li}, {Armitage}, {Carrera},
  {Lyra}, {Rea}, {Yang}, \& {Youdin}}]{Lim2024}
{Lim}, J., {Simon}, J.~B., {Li}, R., {et~al.} 2024, \apj, 969, 130

\bibitem[{{Lim} {et~al.}(2025){Lim}, {Simon}, {Li}, {Carrera}, {Baronett},
  {Youdin}, {Lyra}, \& {Yang}}]{Lim2025}
{Lim}, J., {Simon}, J.~B., {Li}, R., {et~al.} 2025, \apj, 981, 160

\bibitem[{{Lin} \& {Hsu}(2022)}]{Lin2022}
{Lin}, M.-K. \& {Hsu}, C.-Y. 2022, \apj, 926, 14

\bibitem[{{Liu} {et~al.}(2024){Liu}, {Roussel}, {Linz}, {Fang}, {Wolf},
  {Kirchschlager}, {Henning}, {Yang}, {Du}, {Flock}, \& {Wang}}]{Liu2024}
{Liu}, Y., {Roussel}, H., {Linz}, H., {et~al.} 2024, \aap, 692, A148

\bibitem[{{Magnan} {et~al.}(2024){Magnan}, {Heinemann}, \&
  {Latter}}]{Magnan2024}
{Magnan}, N., {Heinemann}, T., \& {Latter}, H.~N. 2024, \mnras, 534, 3944

\bibitem[{{Matthijsse} {et~al.}(2025){Matthijsse}, {Aly}, \&
  {Paardekooper}}]{Matthijsse2025}
{Matthijsse}, J., {Aly}, H., \& {Paardekooper}, S.-J. 2025, \aap, 695, A158

\bibitem[{{McNally} {et~al.}(2021){McNally}, {Lovascio}, \&
  {Paardekooper}}]{McNally2021}
{McNally}, C.~P., {Lovascio}, F., \& {Paardekooper}, S.-J. 2021, \mnras, 502,
  1469

\bibitem[{{Meheut} {et~al.}(2012){Meheut}, {Meliani}, {Varniere}, \&
  {Benz}}]{Meheut2012}
{Meheut}, H., {Meliani}, Z., {Varniere}, P., \& {Benz}, W. 2012, \aap, 545,
  A134

\bibitem[{{Michoulier} {et~al.}(2024{\natexlab{a}}){Michoulier}, {Gonzalez}, \&
  {Price}}]{Michoulier2024_PorosityEvolution_dataset}
{Michoulier}, S., {Gonzalez}, J.-F., \& {Price}, D.~J. 2024{\natexlab{a}},
  dataset, Zenodo, \url{https://doi.org/10.5281/zenodo.12729015}

\bibitem[{{Michoulier} {et~al.}(2024{\natexlab{b}}){Michoulier}, {Gonzalez}, \&
  {Price}}]{Michoulier2024_PorosityEvolution}
{Michoulier}, S., {Gonzalez}, J.-F., \& {Price}, D.~J. 2024{\natexlab{b}},
  \aap, 688, A31

\bibitem[{{Okuzumi} {et~al.}(2012){Okuzumi}, {Tanaka}, {Kobayashi}, \&
  {Wada}}]{Okuzumi2012}
{Okuzumi}, S., {Tanaka}, H., {Kobayashi}, H., \& {Wada}, K. 2012, \apj, 752,
  106

\bibitem[{{Ormel} {et~al.}(2007){Ormel}, {Spaans}, \& {Tielens}}]{Ormel2007}
{Ormel}, C.~W., {Spaans}, M., \& {Tielens}, A.~G.~G.~M. 2007, \aap, 461, 215

\bibitem[{{Paardekooper} {et~al.}(2020){Paardekooper}, {McNally}, \&
  {Lovascio}}]{Paardekooper2020}
{Paardekooper}, S.-J., {McNally}, C.~P., \& {Lovascio}, F. 2020, \mnras, 499,
  4223

\bibitem[{{Paardekooper} \& {Mellema}(2004)}]{Paardekooper2004}
{Paardekooper}, S.~J. \& {Mellema}, G. 2004, \aap, 425, L9

\bibitem[{{Pinte} {et~al.}(2019){Pinte}, {van der Plas}, {M{\'e}nard}, {Price},
  {Christiaens}, {Hill}, {Mentiplay}, {Ginski}, {Choquet}, {Boehler},
  {Duch{\^e}ne}, {Perez}, \& {Casassus}}]{Pinte2019}
{Pinte}, C., {van der Plas}, G., {M{\'e}nard}, F., {et~al.} 2019, Nature
  Astronomy, 3, 1109

\bibitem[{{Pollack} {et~al.}(1996){Pollack}, {Hubickyj}, {Bodenheimer},
  {Lissauer}, {Podolak}, \& {Greenzweig}}]{Pollack1996}
{Pollack}, J.~B., {Hubickyj}, O., {Bodenheimer}, P., {et~al.} 1996, \icarus,
  124, 62

\bibitem[{{Price} {et~al.}(2018){Price}, {Wurster}, {Tricco}, {Nixon},
  {Toupin}, {Pettitt}, {Chan}, {Mentiplay}, {Laibe}, {Glover}, {Dobbs},
  {Nealon}, {Liptai}, {Worpel}, {Bonnerot}, {Dipierro}, {Ballabio}, {Ragusa},
  {Federrath}, {Iaconi}, {Reichardt}, {Forgan}, {Hutchison}, {Constantino},
  {Ayliffe}, {Hirsh}, \& {Lodato}}]{Price2018}
{Price}, D.~J., {Wurster}, J., {Tricco}, T.~S., {et~al.} 2018, \pasa, 35, e031

\bibitem[{Quemener \& Corvellec(2013)}]{Quemener2013}
Quemener, E. \& Corvellec, M. 2013, Linux J., 2013, 3

\bibitem[{{Sch{\"a}fer} {et~al.}(2017){Sch{\"a}fer}, {Yang}, \&
  {Johansen}}]{Schafer2017}
{Sch{\"a}fer}, U., {Yang}, C.-C., \& {Johansen}, A. 2017, \aap, 597, A69

\bibitem[{{Shakura} \& {Sunyaev}(1973)}]{SS73}
{Shakura}, N.~I. \& {Sunyaev}, R.~A. 1973, \aap, 24, 337

\bibitem[{{Squire} \& {Hopkins}(2020)}]{Squire2020}
{Squire}, J. \& {Hopkins}, P.~F. 2020, \mnras, 498, 1239

\bibitem[{{Tazaki} {et~al.}(2023){Tazaki}, {Ginski}, \& {Dominik}}]{Tazaki2023}
{Tazaki}, R., {Ginski}, C., \& {Dominik}, C. 2023, \apjl, 944, L43

\bibitem[{{Tominaga} \& {Tanaka}(2023)}]{Tominaga2023}
{Tominaga}, R.~T. \& {Tanaka}, H. 2023, \apj, 958, 168

\bibitem[{{Vericel} \& {Gonzalez}(2020)}]{Vericel2020}
{Vericel}, A. \& {Gonzalez}, J.-F. 2020, \mnras, 492, 210

\bibitem[{{Vericel} {et~al.}(2021){Vericel}, {Gonzalez}, {Price}, {Laibe}, \&
  {Pinte}}]{Vericel2021}
{Vericel}, A., {Gonzalez}, J.-F., {Price}, D.~J., {Laibe}, G., \& {Pinte}, C.
  2021, \mnras, 507, 2318

\bibitem[{{Weidenschilling}(1977)}]{Weidenschilling1977}
{Weidenschilling}, S.~J. 1977, \mnras, 180, 57

\bibitem[{{Weidenschilling} \& {Cuzzi}(1993)}]{Weidenschilling1993}
{Weidenschilling}, S.~J. \& {Cuzzi}, J.~N. 1993, in Protostars and Planets III,
  ed. E.~H. {Levy} \& J.~I. {Lunine}, 1031

\bibitem[{{Whipple}(1972)}]{Whipple1972}
{Whipple}, F.~L. 1972, in From Plasma to Planet, ed. A.~{Elvius}, 211

\bibitem[{{Yang} {et~al.}(2017){Yang}, {Johansen}, \& {Carrera}}]{Yang2017}
{Yang}, C.-C., {Johansen}, A., \& {Carrera}, D. 2017, \aap, 606, A80

\bibitem[{{Yang} \& {Zhu}(2021)}]{Yang2021}
{Yang}, C.-C. \& {Zhu}, Z. 2021, \mnras, 508, 5538

\bibitem[{{Youdin} \& {Johansen}(2007)}]{Youdin2007}
{Youdin}, A. \& {Johansen}, A. 2007, \apj, 662, 613

\bibitem[{{Youdin} \& {Goodman}(2005)}]{Youdin2005}
{Youdin}, A.~N. \& {Goodman}, J. 2005, \apj, 620, 459

\bibitem[{{Zsom} {et~al.}(2010){Zsom}, {Ormel}, {G{\"u}ttler}, {Blum}, \&
  {Dullemond}}]{Zsom2010}
{Zsom}, A., {Ormel}, C.~W., {G{\"u}ttler}, C., {Blum}, J., \& {Dullemond},
  C.~P. 2010, \aap, 513, A57

\end{thebibliography}

\begin{appendix}

\onecolumn
\nolinenumbers

\section{Summary of results from \citetalias{Michoulier2024_PorosityEvolution}}
\label{App:ResultsSummary}

\begin{figure*}[ht!]
  \centering
  \includegraphics[width=\textwidth,clip]{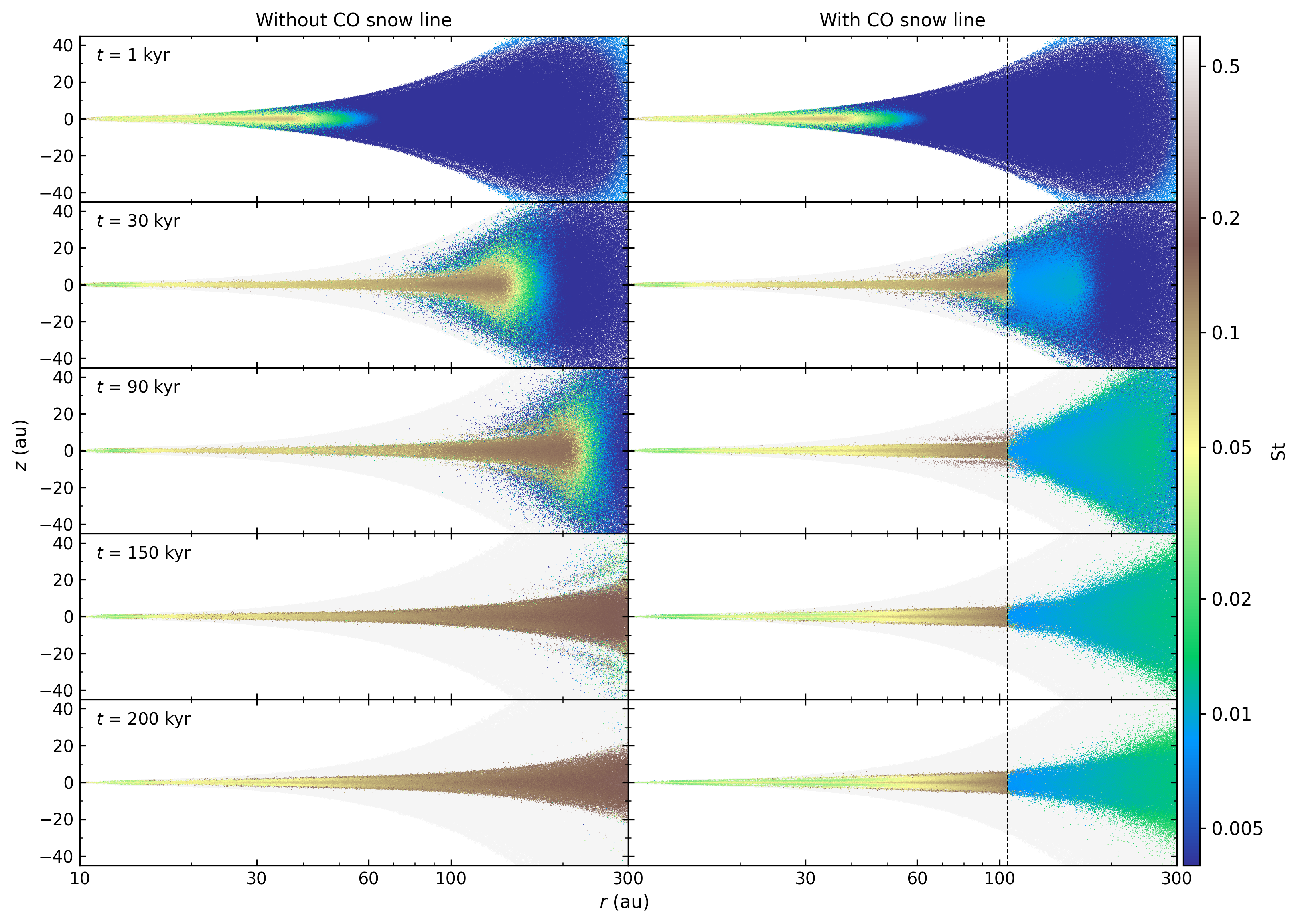}
  \caption{Map of the Stokes number St in the meridian plane $(r,z)$, with the light grey background showing the extent of the gas disc in simulations without (\textit{left column}) and with the CO snow line, marked by the vertical dashed line (\textit{right column}). Time evolution is shown from top to bottom from $t=1$ to 200~kyr.}
  \label{Fig:r-z-St-evol}
\end{figure*}

\begin{figure*}[ht!]
  \centering
  \includegraphics[width=\textwidth,clip]{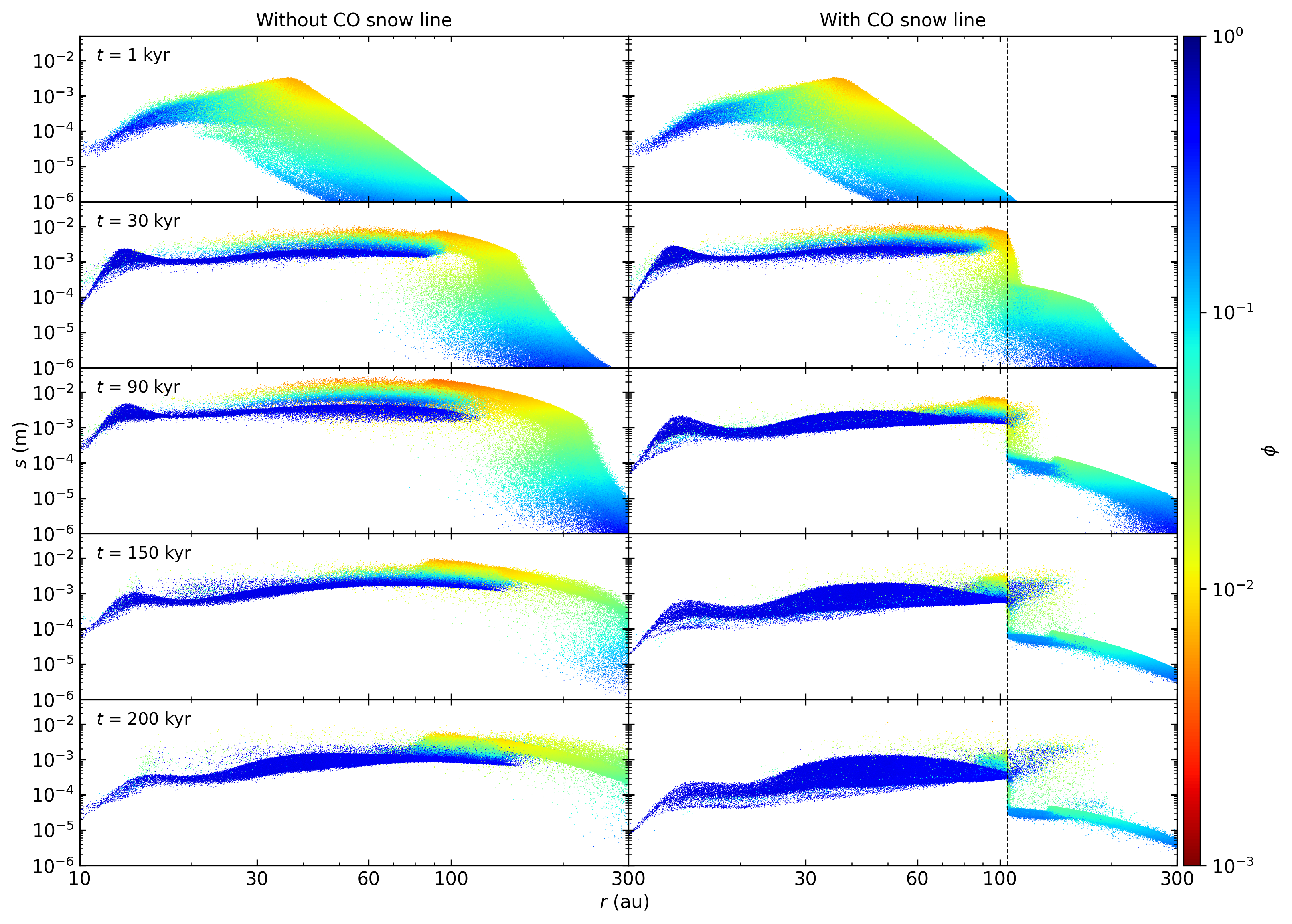}
  \caption{Radial grain size distribution, with colour representing the filling factor $\phi$, in simulations without (\textit{left column}) and with the CO snow line, marked by the vertical dashed line (\textit{right column}). Time evolution is shown from top to bottom from $t=1$ to 200~kyr.}
  \label{Fig:r-s-phi-evol}
\end{figure*}

Figures~\ref{Fig:r-z-St-evol} and \ref{Fig:r-s-phi-evol} recall the results from \citetalias{Michoulier2024_PorosityEvolution} for the two considered simulations (GBFc-Si-a02-Vf20, left and GBFcS-Si-a02-Vf205, right) at relevant snapshots from $t=1$ to 200~kyr. Maps of the grains Stokes number in the meridian plane are displayed in Fig.~\ref{Fig:r-z-St-evol} while Fig.~\ref{Fig:r-s-phi-evol} shows radial grain size distributions coloured by filling factor. Initial dust growth is fast in the inner ($\sim$10 to $\sim$30~au) and intermediate ($\sim$30 to $\sim$100~au) disc regions, with grains reaching a few mm in the latter after 1~kyr, while they are still $\mu$m-sized in the outer disc (exterior to $\sim$100~au, or outside the CO snow line when it is included). After a few tens of kyr, efficient dust settling is seen except in the outer regions. Interior to $\sim$100~au, dust evolution remains very similar in both simulations up to $t=30$~kyr, when grains have reached sizes of a few mm to almost a cm in most of the region. In the intermediate disc, large porous grains with $\phi$ of a few $10^{-3}$ coexist with smaller grains that have been compacted to $\phi\sim0.3$. With $\St\sim0.1$, grains drift towards the inner disc edge, slow down where $\St\sim0.02$ and pile up (see Appendix~\ref{App:TimeEvol}). In the outer disc without CO snow line (left), grains are in the process of growing to mm sizes, while with the CO snow line (right), grains grow less efficiently and to smaller sizes.

Later on, the more fragile grains in the outer disc with the CO snow line stay smaller than 100~$\mu$m, with $\St\lesssim0.02$, and drift slowly, thus starving the intermediate and inner disc of already grown dust, where grains grow more slowly and their St remains $\sim0.1$ at most just inside the CO snow line. Closer in, compaction continues more slowly and St decreases to values $\sim$0.05 or less, slowing down radial drift in the intermediate disc and causing a second pile-up (see Appendix~\ref{App:TimeEvol}). In contrast, without CO snow line grains keep growing in the whole disc and reach their largest sizes of a few cm at 90~kyr. The region of drifting grains with $\St\gtrsim0.1$ extends in the outer disc. At 150~kyr, grains have reached mm sizes and $\St\gtrsim0.2$ out to 300~au, with vigorous compaction now in the 80--150~au region. Similarly to the case with the CO snow line, but with a delay exceeding 100~kyr, compaction then proceeds more slowly in the inner disc, reducing St to $\sim$0.05 at 200~kyr.

\begin{figure*}[ht!]
  \centering
  \includegraphics[width=\textwidth,clip]{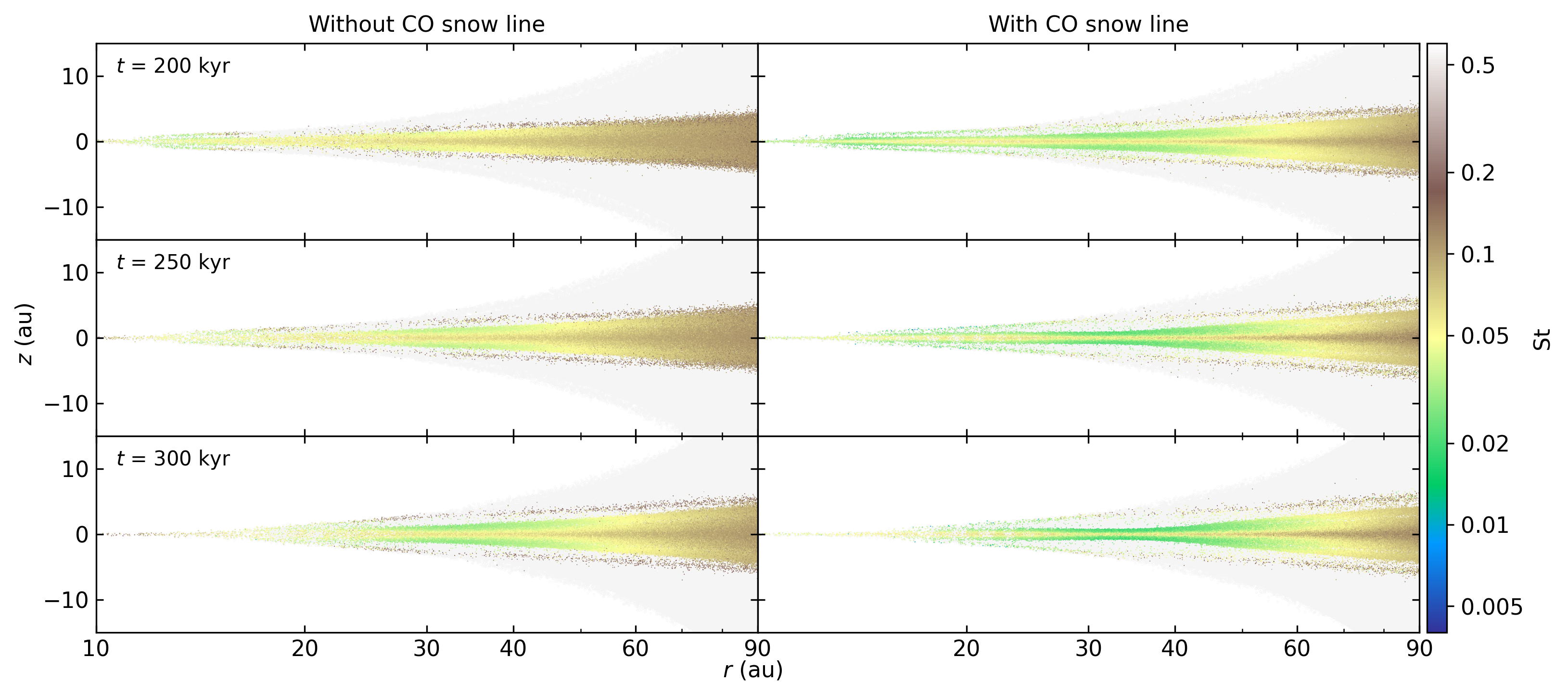}
  \caption{Same as Fig.~\ref{Fig:r-z-St-evol}, zoomed in on the disc interior to 90~au to show the late evolution at $t=200$, 250 and 300~kyr.}
  \label{Fig:r-z-St-zoom-evol}
\end{figure*}

The subsequent evolution is better seen in Fig.~\ref{Fig:r-z-St-zoom-evol}, zooming in on the inner and intermediate disc. Without CO snow line, St keeps decreasing in the intermediate disc, from 0.05--0.1 at 200~kyr around 30--40~au to 0.02--0.05 at 300~kyr. With the CO snow line, the same trend is seen but with values going from 0.02--0.05 at 200~kyr to $<0.02$ at 300~kyr. In both cases, this enhances the dust slowdown and pile-up. Dust settling is also increasing in both simulations, similarly delayed in that without CO snow line.

\vspace*{4ex}

\section{Time evolution of the strong-clumping regions}
\label{App:TimeEvol}

\begin{figure*}[ht!]
  \centering
  \includegraphics[width=\textwidth,clip]{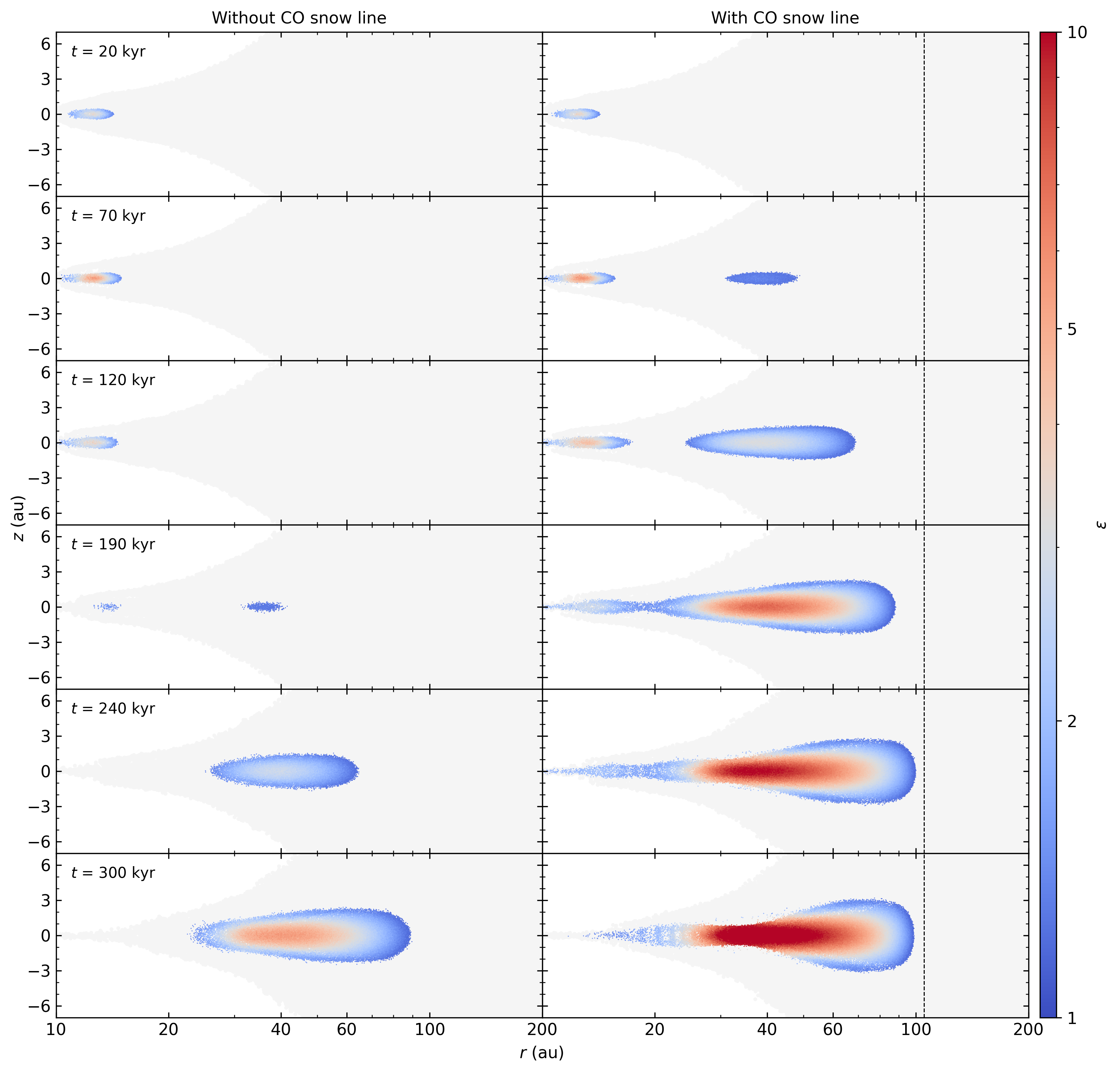}
  \caption{Same as Fig.~\ref{Fig:SI_location}, showing its time evolution from $t=20$ to 300~kyr, from top to bottom.}
  \label{Fig:SI_location-evol}
\end{figure*}

Figure~\ref{Fig:SI_location-evol} shows meridian plane maps of the strong-clumping regions, i.e. where $\varepsilon>\varepsilon_\mathrm{crit}$, at relevant snapshots from \mbox{$t=20$} to 300~kyr. These regions of enhanced dust-to-gas ratio are formed by the accumulation of grains in pile-ups resulting from their evolution. The first of these strong-clumping regions appear near the disc inner edge at 20~kyr, simultaneously in both simulations, as the first pile-up identified in Appendix~\ref{App:ResultsSummary}. At 70~kyr, the second pile-up, forming only when the CO snow line is present (right), leads to the appearance of a new strong-clumping region spanning the 30--50 au region. In both simulations, the inner one reaches high concentrations, with $\varepsilon\gtrsim5$, in a narrow ring at $\sim$11--12~au. Later on, the St gradient become shallower at the disc inner edge, following the grain size gradient evolution (Fig.~\ref{Fig:r-s-phi-evol}), which weakens the inner dust pile-up, faster in the case without CO snow line (left). With the CO snow line, the outer strong-clumping region expands over time and its dust-to-gas ratio increases, reaching $\varepsilon\sim5$ at 150~kyr (Fig.~\ref{Fig:SI_location}) and exceeding 10 after 240~kyr, as dust settling continues. Without CO snow line, the 190~kyr snapshot shows a much weakened inner strong-clumping region (it will disappear by 200~kyr), and the appearance of a new one near 35~au. Its subsequent evolution resembles that of the case with the CO snowline, delayed by over 100~kyr (see Appendix~\ref{App:ResultsSummary}). At 300~kyr, its peak dust-to-gas ratio is similar to that with the CO snow line at 150~kyr.

\section{Mass reservoir for planetesimal formation}
\label{App:MassReservoir}

\begin{figure}[ht!]
  \sidecaption
  \includegraphics[width=12cm,clip]{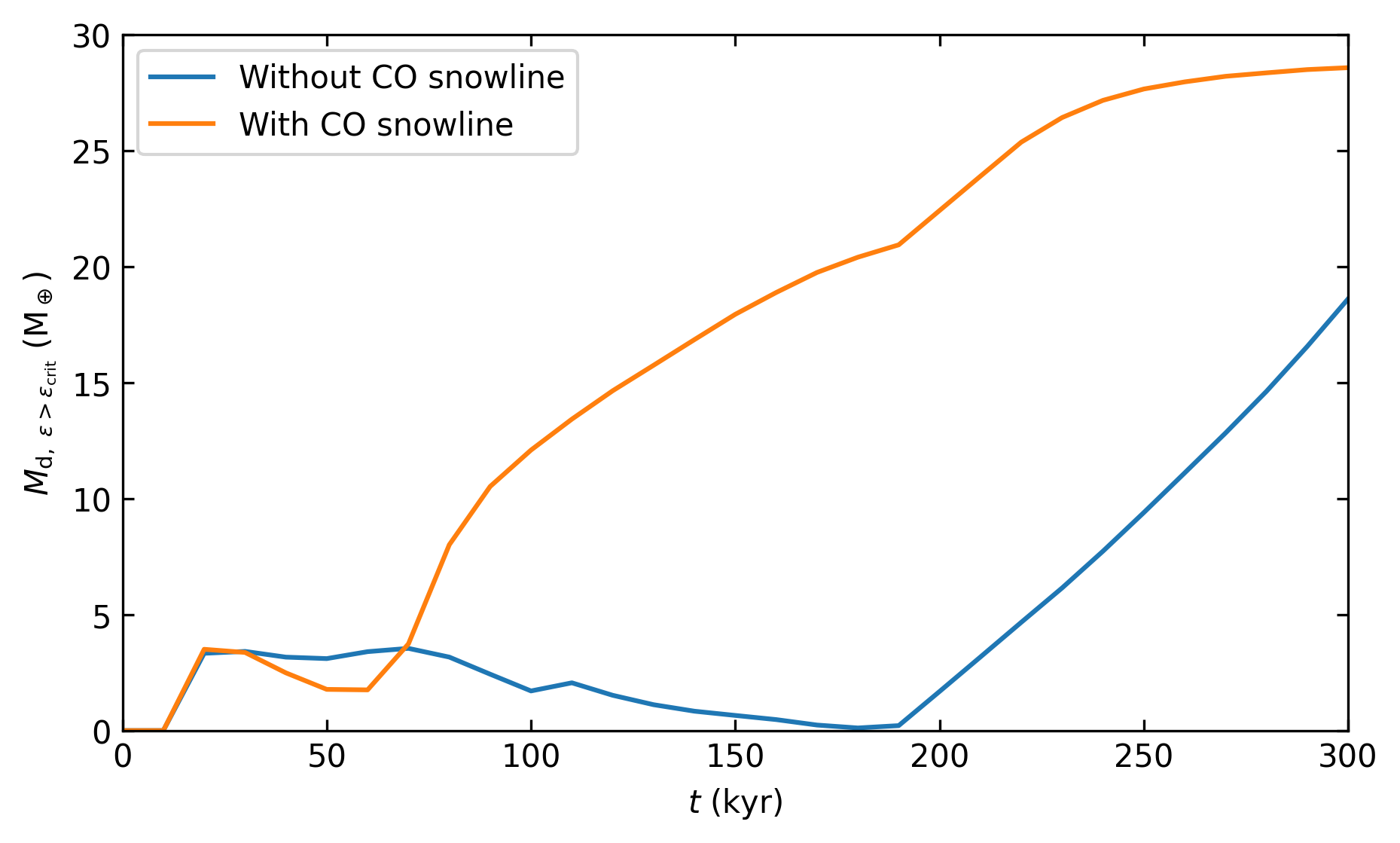}
  \caption{Time evolution of the dust mass in strong-clumping regions $M_{\mathrm{d},\ \varepsilon>\varepsilon_\mathrm{crit}}$, in Earth masses, for the discs without (blue) and with (orange) CO snowline.}
  \label{Fig:Mass_reservoir}
\end{figure}

Our simulations cannot capture the SI (see Sect.~\ref{Sec:Methods}) or the ensuing planetesimal formation. A commonly used recipe to estimate the mass of planetesimals formed by the SI \citep[see, e.g.,][and references therein]{Drazkowska2017} is to assume that a fraction of the dust mass meeting the strong-clumping criterion at a given location is converted into planetesimals during each orbital period. This method is unfortunately impractical to use as a post-treatment of our simulations, since it would mean removing a substantial mass of dust grains, thus affecting the dust mass in strong-clumping regions $M_{\mathrm{d},\ \varepsilon>\varepsilon_\mathrm{crit}}$ at later time steps in a non-self-consistent way. This could be done in new dedicated simulations in a future work. However, $M_{\mathrm{d},\ \varepsilon>\varepsilon_\mathrm{crit}}$ can be used as a proxy for the capacity of a disc to form planetesimals. Ideally, one would like to have a metric for the total mass of planetesimals formed over the disc's lifetime but using a cumulative measure of $M_{\mathrm{d},\ \varepsilon>\varepsilon_\mathrm{crit}}$ is again impractical as a fraction of dust grains would be counted several times. Despite those caveats, $M_{\mathrm{d},\ \varepsilon>\varepsilon_\mathrm{crit}}$ can still be useful to compare the discs without and with the CO snow line. Its time evolution is shown in Fig.~\ref{Fig:Mass_reservoir}. With the CO snowline, $M_{\mathrm{d},\ \varepsilon>\varepsilon_\mathrm{crit}}$ increases rapidly after 70~kyr, exceeds 10~M$_\oplus$ at 90~kyr and reaches 28~M$_\oplus$ by 300~kyr. In contrast, without CO snowline, the fast increase of $M_{\mathrm{d},\ \varepsilon>\varepsilon_\mathrm{crit}}$ only starts at 190~kyr, and it exceeds 10~M$_\oplus$ only after 250~kyr. The CO snowline thus appears to promote the early formation of planetesimals.

\end{appendix}
\end{document}